\documentclass[preprint2]{aastex}

\slugcomment{To appear in the Astrophysical Journal}

\shorttitle{Large-Scale Periodic Variability in  WR\,1}
\shortauthors{Chen\'e \& St-Louis}

\begin{document}

\title{Large-Scale Periodic Variability of the
    Wind of the Wolf-Rayet Star WR\,1 (HD 4004)}

\author{A.-N. Chen\'e\altaffilmark{*}$^,$ \altaffilmark{**}}
\affil{Canadian Gemini Office, HIA/NRC of Canada, \\
5071, West Saanich Road, Victoria (BC), V9E 2E7, Canada}
\email{andre-nicolas.chene@nrc-cnrc.gc.ca}

\and

\author{N. St-Louis}
\affil{D\'{e}partement de Physique and CRAQ,
Universit\'{e} de Montr\'{e}al, C.P. 6128,
Succ. Centre-Ville, Montr\'{e}al, Qu\'{e}bec, H3C 3J7, Canada}
\email{stlouis@astro.umontreal.ca}

\vspace{1cm}

\altaffiltext{*}{Guest investigator, Dominion Astrophysical
Observatory, Herzberg Institute of Astrophysics, National Research
Council of Canada.}

\altaffiltext{**}{Based on observations obtained at the
Canada-France-Hawaii Telescope (CFHT) which is operated by the
National Research Council of Canada, the Institut National des
Sciences de l'Univers of the Centre National de Recherche Scientifique
of France, and the University of Hawaii. Also based on observations
obtained at the Observatoire du Mont M\'egantic with is operated by
the Centre de Recherche en Astrophysique du Qu\'ebec and the
Observatoire de Haute-Provence which is operated by the Institut
National des Sciences de l'Univers of the Centre National de la
Recherche Scientifique of France.}

\begin{abstract}
We present the results of an intensive photometric and spectroscopic
monitoring campaign of the WN4 Wolf-Rayet (WR) star WR\,1=HD~4004. Our
broadband $V$ photometry covering a timespan of 91~days shows
variability with a period of P=16.9$^{+0.6}_{-0.3}$~days. The same
period is also found in our spectral data.

The light-curve is non-sinusoidal with hints of a gradual change in
its shape as a function of time. The photometric variations
nevertheless remain coherent over several cycles and we estimate that
the coherence timescale of the light-curve is of the order of 60
days. The spectroscopy shows large-scale line-profile variability
which can be interpreted as excess emission peaks moving from one side
of the profile to the other on a timescale of several days.

Although we cannot unequivocally exclude the unlikely possibility that
WR\,1 is a binary, we propose that the nature of the variability we
have found strongly suggests that it is due to the presence in the
wind of the WR star of large-scale structures, most likely Co-rotating
Interaction Regions (CIRs), which are predicted to arise in inherently
unstable radiatively driven winds when they are perturbed at their
base. We also suggest that variability observed in WR\,6, WR\,134 and
WR\,137 is of the same nature. Finally, assuming that the period of
CIRs is related to the rotational period, we estimate the rotation
rate of the four stars for which sufficient monitoring has been
carried out; i.e. v$_{rot}$=6.5, 40, 70 and 275~km~s$^{-1}$ for WR\,1,
WR\,6, WR\,134 and WR\,137, respectively.
\end{abstract}


\section{Introduction}

The winds of Wolf-Rayet (WR) stars are well-known to be non-uniform on
small physical scales. Because of the inherently unstable nature of
radiatively-driven hot stellar winds \citep{OCR}, instabilities, which
reveal themselves in the spectrum as narrow excess emission peaks
superposed on the broad emission lines, appear stochastically,
propagate in the wind and disappear after several hours
\citep[e.g. ][]{Moffat88}. This clumpiness leads to emission lines
that are variable at a level of up to $\sim$~5~\% of the line flux
\citep{Stlouis}. These changes are, of course, random and no
periodicity is expected.

In certain cases, WR stars also display large-scale line-profile
variability (lpv). For some stars this spectroscopic variability has
been shown to be periodic and to originate from a different physical
processes. Of course, known massive WR+O binaries produce clear
periodic radial-velocity (RV) variations from the orbital motion of
the stars. It has also been realized in the past two decades that in
such systems the winds from the stars collide, forming a shock cone
that wraps around the star with the smaller momentum flux (generally
the O star). This interaction induces kinematically-characteristic
variations in the line {\it profiles} that are also periodic.

However, large-scale lpv has also been found in WR stars that are not
known to be binaries. The two most studied cases, WR\,6 (WN4) and
WR\,134 (WN6), have been monitored intensively in photometry and
spectroscopy by \citet{Morel97,Morel98,Morel99a}. In both cases, these
authors have observed periodic large-scale lpv and a complex
light-curve with the same periodicity. To that list of two stars, we
may also add WR\,137, a WC7pd star in a long-period binary
system. This star shows large-scale lpv with a period of 1.2~d that
is unlikely to be related to its O9 companion nor to the wind-wind
collision zone \citep{Lef}. This behaviour is most likely explained by
the presence of large-scale density structures in the wind, such as
co-rotating interaction regions (CIRs) \citep{Cranmer96,Fullerton97},
but the possibility of the presence of a compact companion or of a
low-mass main-sequence star cannot be completely excluded.

If the large-scale lpv and the photometric variability of single WR
stars can be associated with CIRs, it may imply that the period of
these variabilities corresponds directly to the rotational period of
the star. Indeed, \citet{Cranmer96} propose that ``spots'' fixed to
the stellar surface, caused either by pulsations or magnetic field
activity, are at the origin of the CIRs. Hence, once the radius at
which the CIR originates is known, we can determine the rotational
velocity of the star at that point. It follows that the rotation rate
of at least some WR stars could in principle be determined by carrying
out a systematic investigation of the variability of all single WR
stars. As a first step, \citet{Stlouis} and Chen\'e \& St-Louis (in
prep) set out to identify new candidates for CIR-type variability by
conducting a survey of all apparently single Galactic WR stars
brighter than $v\sim$12.5$\rm ^{th}$ magnitude. For each star in their
sample, they obtained 4--5 spectra which allowed them to establish a
list of WR stars showing large-scale lpv. The next step is to observe
intensively each of the candidates in order to verify and determine
the periodicity.

In the above-mentioned survey, WR\,1 was one of the most striking
cases of large-scale lpv with changes reaching 8--10\% of the line
flux and easily distinguishable large scale subpeaks superposed on the
broad wind emission profiles. Consequently, it was the first WR star
on which we concentrated our efforts. Several previous studies
claiming unreproducible periods for the variability of this star have
been summarized in \citet{Morel99b}. Later, \citet{Nieda,Niedb}
demonstrated that the spectrum of WR\,1 varied greatly from
night-to-night reaching 50\% in the equivalent width of the He{\sc
i}$\lambda$5876 line. On the other hand, they found that line-profile
changes during one night are much smaller with a typical scatter of
0.5\,\AA\, in equivalent width for the He{\sc ii}$\lambda$5411, C{\sc
iv}$\lambda$5808 and He{\sc i}$\lambda$5876 lines. The changes,
however, were found to be systematic during the course of an entire
night and reach a total of 3-4\,\AA. Moreover, the search for periods
smaller than 2 days failed and only indications of long-term
variability could be suggested. The investigation of photometric
variability over more than 16~days carried out by \citet{Morel99b} did
not lead to the identification of a period. Consequently, the period
search must be done using data taken over a time range of at least
twice as long. More recently, \citet{Flor} claimed a period of
P=7.684~days in lpv of this star which they attribute to the ejecta of
streams or jets from the stellar surface.

In this paper, we present the results of an intensive monitoring
campaign of WR\,1 extending over several weeks in photometry and
spectroscopy with the aim of determining the nature of the variability
and, eventually, if found to be associated with CIRs as suspected from
our survey observations, the rotation period of WR\,1. In
Section~\ref{obs}, we present our photometric and spectroscopic
observations and the data reduction procedures. In
Sections~\ref{resPhot} and \ref{ressp} we describe our results, and in
Section~\ref{dis} we discuss the possible interpretations. Finally,
our conclusions are presented in Section~\ref{con}.

\section{Observations}\label{obs}

\subsection{Photometry}\label{obsphot}

We monitored WR\,1 in broadband $V$ using CCD-imagery at the 0.81m
Tenagra Observatory ltd. from 20 November 2003 to 18 February
2004. During this period covering 91 nights, three frames were
obtained in succession every clear night with an exposure time of
60s. Three additional frames with a 45s integration time were obtained
at an airmass of $\sim$1.45 during the first 43 nights and three
others with a 30s integration time at an airmass of $\sim$1.65 during
the first 11 nights. Due to problems with the detector in
December 2003--January 2004, which resulted in a gap in our dataset of
almost 20~days, 33~frames were unusable. Also, 87 others were rejected
due to bad seeing and/or low transparency of the sky, giving finally a
total of 231 usable frames. In order to correct for the differential
refraction of the atmosphere, we observed three frames in broadband B
for each of two very clear nights at Tenagra Observatory ltd.,
i.e. the 6$^{th}$ and the 17$^{th}$ of December 2003, for a total of 6
frames.

All images were uniformly reduced using standard procedures carried
out with routines written in Interactive Data Language (IDL). After
the bias, dark and flat-field treatments, we performed aperture
photometry using the {\sc aper} Astrolib routine in IDL on all stars
present in the field of WR\,1. We adopted an aperture size equal to
twice the FWHM of the point spread function (PSF) and an annulus of
sky was selected with inner and outer radii of respectively 4 and 8
times the FWHM of the PSF.

Once the photometry was obtained for all frames, we then applied a
correction for the differential refraction of the atmosphere. To do
so, we needed to know how the flux of a star varies with airmass and
according to the slope of its spectrum. For each star in our field of
view, a relative ($B-V$) color was obtained by subtracting the average
fluxes measured in the $B$ and $V$-band frames observed at the same
airmass. We then obtained the slope {\it M} of the linear relation
describing the variability of the flux as a function of airmass for
each star during the best clear nights. Although a certain number of
stars can present intrinsic flux variations within a single night, we
can assume that it is not the case for most of them. Using a robust
least absolute deviation method, we fitted {\it M}$_{(B-V)}$, the
slope of the flux-airmass relation appropriate for a given (B-V)
color, and obtained a linear expression for the correction that must
be applied to the flux as a function of airmass. The final corrections
are contained in a range of 0.1 to 1~\%.

The final light-curves were obtained by subtracting the magnitude of
WR\,1 to that of two comparison stars (see
Figure~\ref{photWR1t}). Since no star with a magnitude similar to that
of WR\,1 was available in the field of view, we decided to average the
brightest stars showing small light deviations. Ten stars were
selected and divided in two groups to create two comparison stars of
similar magnitudes. The accuracy of the final light-curve of WR\,1
varied during our observing period. We have thus estimated a typical
error for each day as follows~: for a time interval of 10~days
centered on a given day of observation, we have calculated the
standard deviation ($\sigma$) of the c1-c2 light curve. All values are
shown in the bottom panel of Figure~\ref{photWR1t}. For the entire
observing period, the standard deviation is $\sigma$=0.012 mag.

\begin{table}[htbp]
    \caption{Spectroscopic observing campaigns}
    \begin{tabular}{cclcccc}

      \hline
      Run \# & Telescope & \multicolumn{1}{c}{Dates} & $\lambda$ Coverage & $\Delta\lambda$ & No. of sp. & SNR \\
            &            & \multicolumn{1}{c}{(UT)}  &        \AA         &    (3 pix.)     &            &     \\
      \hline
      1 & OHP 1.52m & 2003/07/07-14 & 5230-6140 & 1.4 \AA & 30 & 135 \\
      2 & CFHT & 2003/08/08-13 & 4380-6240 & 1.5 \AA & 33 & 185 \\
      3 & OMM & 2003/08/22-24  & 4900-6020 & 1.6 \AA & 9 & 130  \\
      4 & OMM & 2004/08/21 -- & 4310-6440 & 1.6 \AA & 170 & 145 \\
        &     & \multicolumn{1}{c}{2004/09/06} & &  &     &     \\
      5 & DAO 1.85m & 2004/09/18 -- & 5140-5980 & 1.4 \AA & 84 & 90 \\
        &           & \multicolumn{1}{c}{2004/10/13} & & & &    \\
      \hline
    \end{tabular}
    \label{resObs}
\end{table}

\subsection{Spectroscopy}

We collected a total of 326 spectra of WR\,1 during five dedicated
runs distributed in July-August 2003 and August-October 2004 with the
1.52m telescope of the Obervatoire de Haute-Provence (OHP), the 3.6m
Canada-France-Hawaii Telescope (CFHT), the 1.6m telescope of the
Observatoire du Mont M\'egantic (OMM) and the 1.85m telescope of the
Dominion Astronomical Observatory (DAO). The details of these
observing runs are summarized in Table~\ref{resObs}. We list the run
number, the telescope used, the dates of the observations, the
wavelength coverage, the spectral resolution, the number of spectra
obtained and the average signal-to-noise ratio (SNR). Unfortunately,
no spectra were obtained simultaneously with our photometry described
in the previous section.

The bias subtraction, flat-fielding, spectrum extraction, sky
subtraction and wavelength calibration of all spectra were executed in
the usual way using the {\sc iraf}\footnote{{\sc iraf} is distributed
by the National Optical Astronomy Observatories (NOAO), which is
operated by the Association of Universities for Research in Astronomy,
Inc. (AURA) under cooperative agreement with the National Science
Foundation (NSF).} software. Calibration-lamp spectra were taken every
30-40 mins depending on the run. The accuracy of the wavelength
calibration estimated by measuring the wavelength of 10 lamp emission
lines is variable depending on the instrument used. It is 0.005\,\AA\,
for OHP, 0.2\,\AA\, for $\lambda <$~5200\,\AA\, and 0.02\,\AA\, for
$\lambda >$~5200\,\AA\, for CFHT, 0.01\,\AA\, for OMM and 0.04\,\AA\,
for DAO. Special care was taken for the normalization of the
spectra. First, a mean was made for each run. Then each spectrum of a
run was divideded by the run mean and the ratio fitted with a low
order Legendre polynomial (between 4$^{th}$ and 8$^{th}$ order). The
original individual spectrum was divided by this fit, and was
therefore at the same level as the run mean. When this procedure was
done for each run, the run means were then put at the same level by
using the same procedure as described above. This allowed us to put
all individual spectra at the same level. Then, we combined all run
means into a global, high quality mean, which was then fitted in
selected pseudo-continuum regions, i.e. wavelength regions where large
emission-lines do not dominate. More specifically, these regions are~:
4376.0 -- 4381.0\,\AA, 4473.0 -- 4477.0\,\AA, 4798.0 -- 4816.0\,\AA,
5002.0 -- 5147.0\,\AA, 5349.0 -- 5362.0\,\AA, 5558.0 -- 5663.0\,\AA,
5956.0 -- 5992.0\,\AA\ and 6295.0 -- 6321.0\,\AA. These regions are
shown in Figure~\ref{tvs}, which we will discuss later in
Section~\ref{ressp}. Finally, the fitted continuum function is applied
to each individual spectrum. The error on the continuum normalization
measured as the standard deviation of individual spectra around the
continuum function is typically of 0.5\%.

\section{Photometric Variations}\label{resPhot}
\begin{figure}[ht]
 \plotone{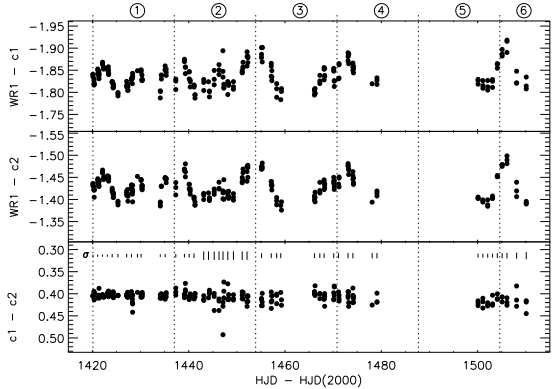}
 \caption{Light-curve of WR\,1. The {\it top} and {\it middle panels}
 show the difference between the magnitude of WR\,1 and that of the
 ``artificial'' comparison stars c1 and c2 respectively as a function
 of time. Both c1 and c2 were constructed from the average of five
 bright stars. The 6 cycles of the 16.9-day period covered by the
 photometric dataset are delimited. The {\it bottom panel} shows the
 difference between the magnitude of c1 and c2 as a function of
 time. A series of vertical lines show the typical error as a function
 of time (see text for details).}
 \label{photWR1t}
\end{figure}

The light-curve of WR\,1 is plotted as a function of the Heliocentric
Julian Date (HJD) in Figure~\ref{photWR1t}. The flux from the star
alternatively increases and decreases with an amplitude that varies
between $\rm \Delta V$=0.06-0.12 mag within 5-7~days ($\rm
\sigma(c1-c2)\sim0.012$ mag), suggesting that the changes are periodic
with a period of at least 5~days. The variability within a single
night is small and typically reaches an amplitude not higher than
2$\sigma$. Since only slow variability is observed in photometry and
spectroscopy within a single night \citep{Morel99b,Nieda}, we binned
all the data obtained during a given night to increase the SNR.

\begin{figure}[ht]
 \plotone{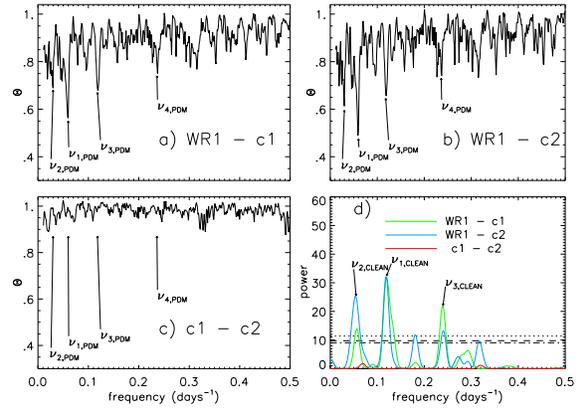}
 \caption{Periodograms: a) $\Theta$-spectrum obtained from the PDM
    analysis of the WR\,1 - c1 curve. The four frequencies
    $\nu_{1,PDM}$, $\nu_{2,PDM}$, $\nu_{3,PDM}$ and $\nu_{4,PDM}$ are
    indicated with an arrow. b) Same as a) for WR~1-c2. c) Same as a)
    for c1-c2. d) C{\sc lean}ed power spectrum of WR~1-c1 (green),
    WR~1-c2 (bleu) and c1-c2 (red). Three levels of confidence at
    99\%, 95\% and 90\% are plotted in black.  The three frequencies
    $\nu_{1,CLEAN}$, $\nu_{2,CLEAN}$ and $\nu_{3,CLEAN}$ are indicated
    with an arrow.}
 \label{periodo}
\end{figure}

We performed a period search using two independent methods. The first
is a phase-dispersion minimization (PDM) algorithm \citep{pdm1} which
is well suited to cases in which only a small number of observations
are available over a limited period of time, especially if the signal
is highly non-sinusoidal. The method consists in folding the data
points in phase using different trial periods and dividing the
resulting curves in a predetermined number of phase bins. For each
trial period we can define $S_j^2$, the variance of bin $j$, $S^2$,
the variance of all $S_j^2$ and finally $\Theta=S^2/ \sigma^2$, where
$\sigma^2$ is the variance of all data points. The other method used
is the {\sc clean} algorithm \citep{Scargle1,clean1} which has the
advantage of taking into account the unevenness of the data sampling,
since it ``cleans'' the discrete Fourier spectrum with a window
function. Both, the $\Theta$ spectra from the PDM method and the power
spectra from the {\sc clean} method are shown in Figure~\ref{periodo}
for WR~1-c1, WR~1-c2 and c1-c2.

The total time elapsed between the first and last point of the
light-curve is 91~days and, as mentioned above, we combined the data
to obtain one measurement per night. Therefore, the search was
performed over a range of frequencies from 2/91 to 0.5 day$^{-1}$ with
a step of 0.00111 day$^{-1}$, respecting the Nyquist criterion. When
using the PDM method for (WR~1-c1) and (WR~1-c2), one frequency
($\nu_{1,PDM}$) is found to be predominant over all
others. Interestingly, when only the first 60 nights are included in
the period search, $\nu_{1,PDM}$ becomes even more significant, and 3
other frequencies can be distinguished from the noise in the
periodogram. The strongest detection is at
$\nu_{1,PDM}$=0.059$^{+0.001}_{-0.002}$~day$^{-1}$ and the three
others are at $\nu_{2,PDM}$=0.028$^{+0.001}_{-0.002}$,
$\nu_{3,PDM}$=0.118$^{+0.001}_{-0.002}$ and
$\nu_{4,PDM}$=0.236$^{+0.001}_{-0.002}$~day$^{-1}$. Note that the
peaks at $\nu_{2,PDM}$, $\nu_{3,PDM}$ and $\nu_{4,PDM}$ reach
comparable levels and are much smaller than the peak at
$\nu_{1,PDM}$. The periodogram of (c1-c2) shows no signal at all,
indicating that there is no periodic variability present in the
light-curve of either of the comparison stars. The power spectra
obtained using the {\sc clean} algorithm for both the (WR~1-c1) and
(WR~1-c2) light-curves show only three peaks that reach values higher
than the 99\% confidence level threshold and that are absent in the
comparison light-curve (c1-c2). These peaks are at
$\nu_{1,CLEAN}$=0.119$\pm$0.007 ($\sim \nu_{3,PDM}$),
$\nu_{2,CLEAN}$=0.057$\pm$0.007 ($\sim \nu_{1,PDM}$) and
$\nu_{3,CLEAN}$=0.240$\pm$0.007 day$^{-1}$ ($\sim \nu_{4,PDM}$).

\begin{figure}[ht]
 \plotone{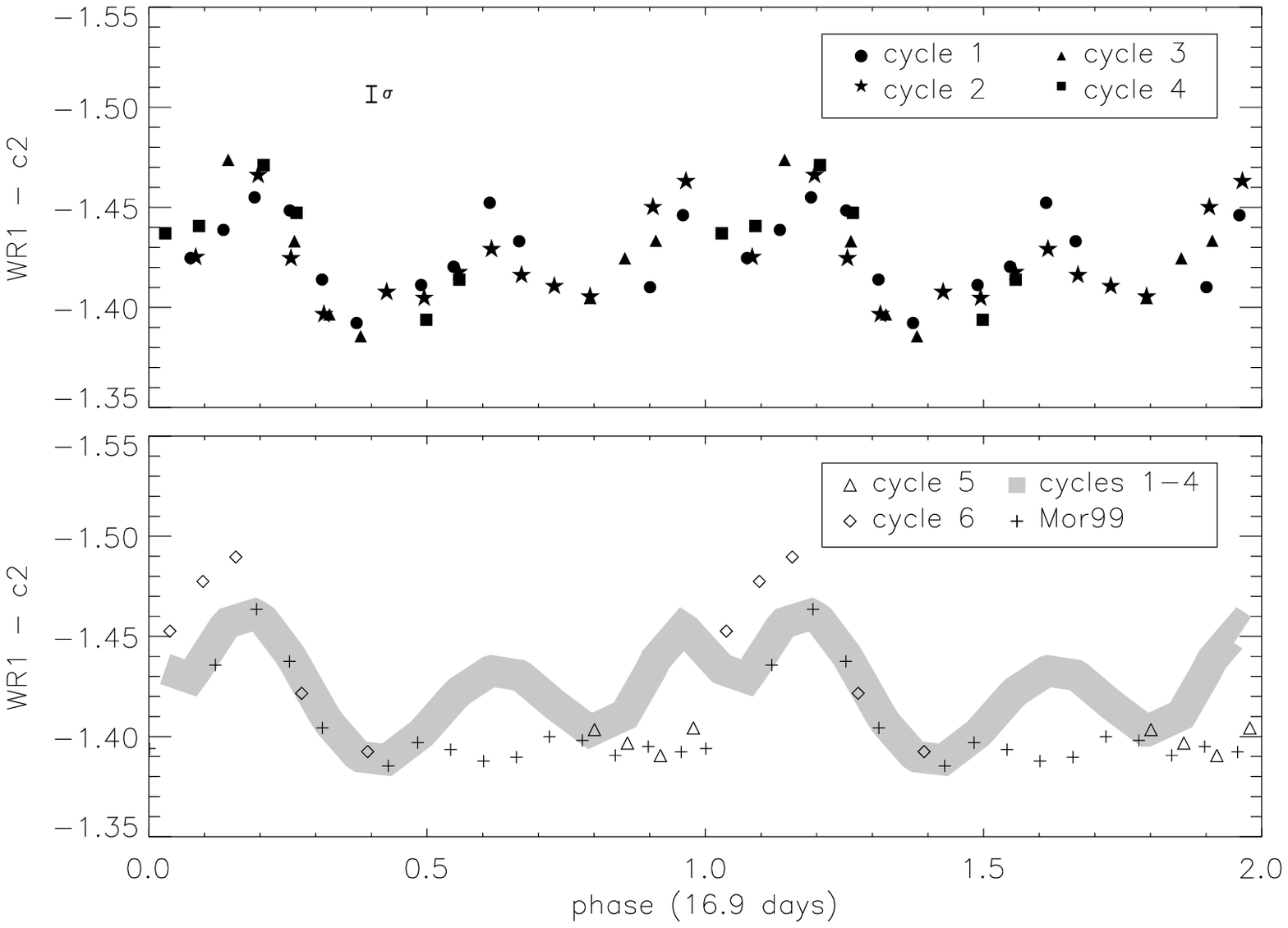}
 \plotone{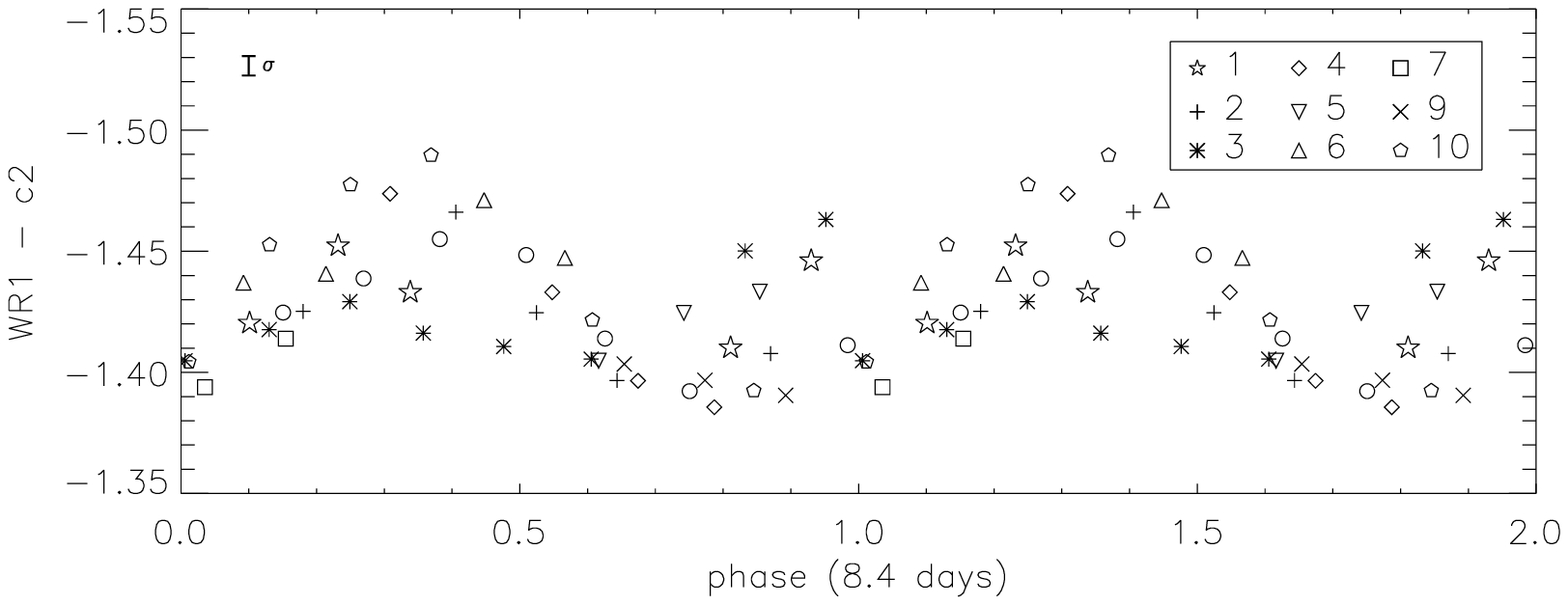}
 \caption{{\it Top panel}: V-band light-curve of WR\,1 folded with the
 16.9~days period. Phase 0 is arbitrary. The first cycle is
 represented by circles, the second by stars, the third by triangles
 and the fourth by squares. {\it Middle panel}: The light-curve from
 the top panel is reproduced here and drawn with a continuous thick,
 gray line. Overplotted to it are the data from cycles 5 (triangles)
 and 6 (diamonds). The narrowband $v$ light-curve obtained by
 \citet{Morel99b} is also plotted with + signs. {\it Bottom panel}:
 The same data folded with a 8.4-day period. The symbols represent the
 9 different cycles of the 8.4-day period.}
 \label{photWR1p}
\end{figure}

The two main frequencies identified using the two different methods
($\nu$=0.059, 0.119 day$^{-1}$) can be linked by an integer number. It
is therefore clear that one is an harmonic of the other. According to
the PDM analysis, the frequency with the strongest signal is
$\nu_{1,PDM}$=0.059$^{+0.001}_{-0.002}$~day$^{-1}$ (period of
16.9$^{+0.6}_{-0.3}$~days), but according to the {\sc clean}ed
spectrum, it is $\nu_{1,CLEAN}$=0.119$\pm$0.007 day$^{-1}$ (period of
8.4$\pm$0.5~days). In an attempt to settle which among these two
possible periods for WR\,1 is the correct one, we folded the
light-curve in phase using the two different frequencies. The
frequency that produced the smallest dispersion and the clearest curve
is $\nu_{1,PDM}$. This result is presented in Figure~\ref{photWR1p}
where each 16.9-day cycle of (WR~1-c2) is plotted using a different
symbol. The 6 cycles of 16.9~days covered by our photometric dataset
are delimited in Figure~\ref{photWR1t}. The 16.9-day folded
light-curve for cycles 1 to 4 shown in the top panel of
Figure~\ref{photWR1p} shows basically three bumps; one broad and small
centered at $\phi\sim$0.6 and two stronger ones spaced closer in phase
at $\phi\sim$0.95 and 0.15. The bottom panel shows the same data
folded with a 8.4-day period. Clearly the scatter is much larger than
when folded with the shorter period. A careful examination of
Figure~\ref{photWR1t} also shows that there is no clear pattern
repeating in any successive 8.4~days, which supports this result. The
fifth and sixth cycles demonstrates the epoch-dependancy of the light
curve. Indeed, the curve shown in the middle panel of
Figure~\ref{photWR1p} shows the disappearance of the second bump near
phase 0.95. Indeed, the third bump seems significantly higher and
wider than during the previous cycles and starts from a lower level,
but the star seems to have skipped the previous increase and decrease
in flux, as if the two nearby peaks had merged. We have no data at
phase 0.6 for these cycles so we cannot tell if the bump at that
location is still present. To show the difference between the shapes
of the light curves, we overplot in the middle panel showing cycles 5
and 6, the cycle 1 to 4 light curve as a thick grey curve which
roughly emcompasses all data points. This epoch dependency of the
light curve of WR\,1 shows that a period can only be found when
observations are taken consecutively, during a period of time for
which the physical conditions of the region where the continuum flux
originates do not change. This means that no data taken after a
certain coherence timescale can be added to the sequence in order for
a period search to be successful. Indeed, even if the period remains
the same, the amplitude and the shape of the light-curve will
change. We cannot determine with accuracy the time of coherence for
WR\,1, but according to our data, it seems to be at least 4 cycles,
i.e. $\sim$60~days.

A light-curve was obtained by \citet{Morel99b}. During their
observations extending over a period of 16~days, the authors found
only one bump lasting 5-days in width, followed by a period of 11~days
during which the curve was flat. In the context of an epoch-dendendant
light curve shape, it is interesting to note that their curve is not
inconsistent with our choice of adopting a period of P$\sim$16.9~days
as the single bump observed by \citet{Morel99b} is similar in height
and width to the bump obtained in the present data during cycle 6 near
phase 0.15. This is shown in the middle panel of Figure~\ref{photWR1p}
where we have arbitrary shifted the curve in phase. On the other hand,
the fact that the star remained constant during 11~days does not bring
support to a shorter 8.4-day period.

If the period of 16.9~days is the real one, why did the {\sc clean}
analysis yield a period with half this value? The answer may reside in
the way the algorithm works. Indeed, the discrete Fourier analysis
looks for a sinusoidal signal. Now, the light-curve of WR\,1 we
observed can be interpreted as two nearly similar sinusoidal curves
per cycle. This is the reason why the {\sc clean}ed spectrum has more
power in the peak at $\nu_{1,PDM}$/2. However, when this frequency is
used to fold the light-curve in phase, the smaller bump is superposed
on the bigger one, and a large dispersion around the curve is
found. This larger dispersion explains why the PDM analysis considered
the $\nu_{1,PDM}$/2 result of lower interest. From this and more
importantly when taking into account the data of \citet{Morel99b} we
conclude that the most probable interpretation is that the period of
16.9$^{+0.6}_{-0.3}$~days, obtained using both methods, is the real
period and that the other detections are simple harmonics or a consequence
of the fact that the light-curve is not a simple sinusoide. Note that
the 7.684-d period claimed by \citet{Flor} has been obtained from
spectra taken during two different runs of 5 and 8 nights separated by
$\sim$6 years. Due to an insufficient time coverage and to the
complexity of the variability pattern, they have detected only a
fraction of the real period.

\section{Spectroscopy}\label{ressp}

The photometry discussed in the previous section provides information
on the variability of the continuum of the star since for this star
only 10\% of the flux in broadband $V$ comes from emission lines. This
continuum originates in deep regions of the stellar wind. In order to
probe the wind at different radii, we must study spectral variability,
because the different WR emission-lines are formed in regions of
different velocity regimes (i.e. different radii), depending on the
ionisation potential of the given elements
\citep[e.g. ][]{Kuh73,Hil87}. From spectral variability, we hope to
obtain clues as to the origin of the period found in photometry.

\subsection{Level of Variability}

\begin{figure}[ht]
    \plotone{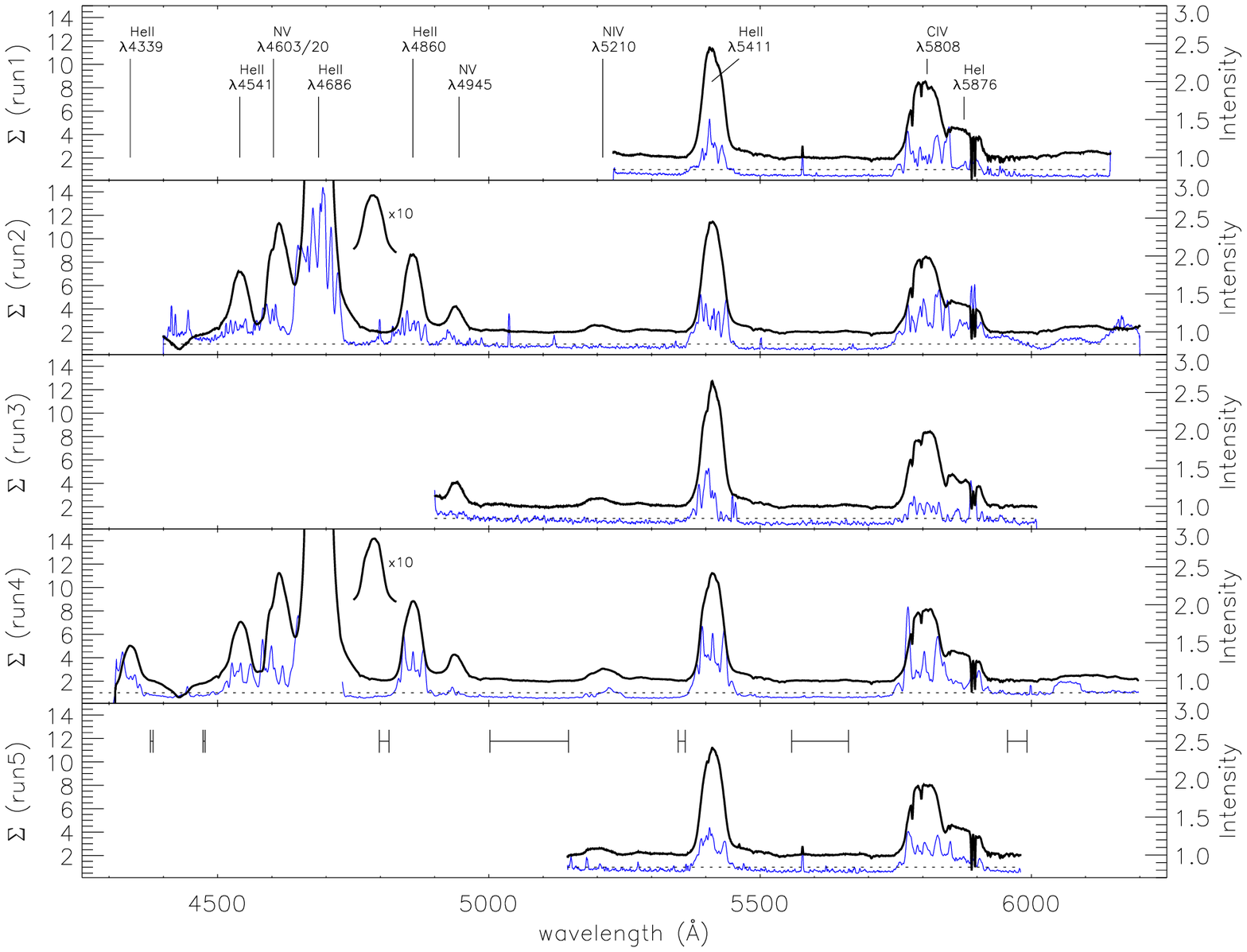}
    \plotone{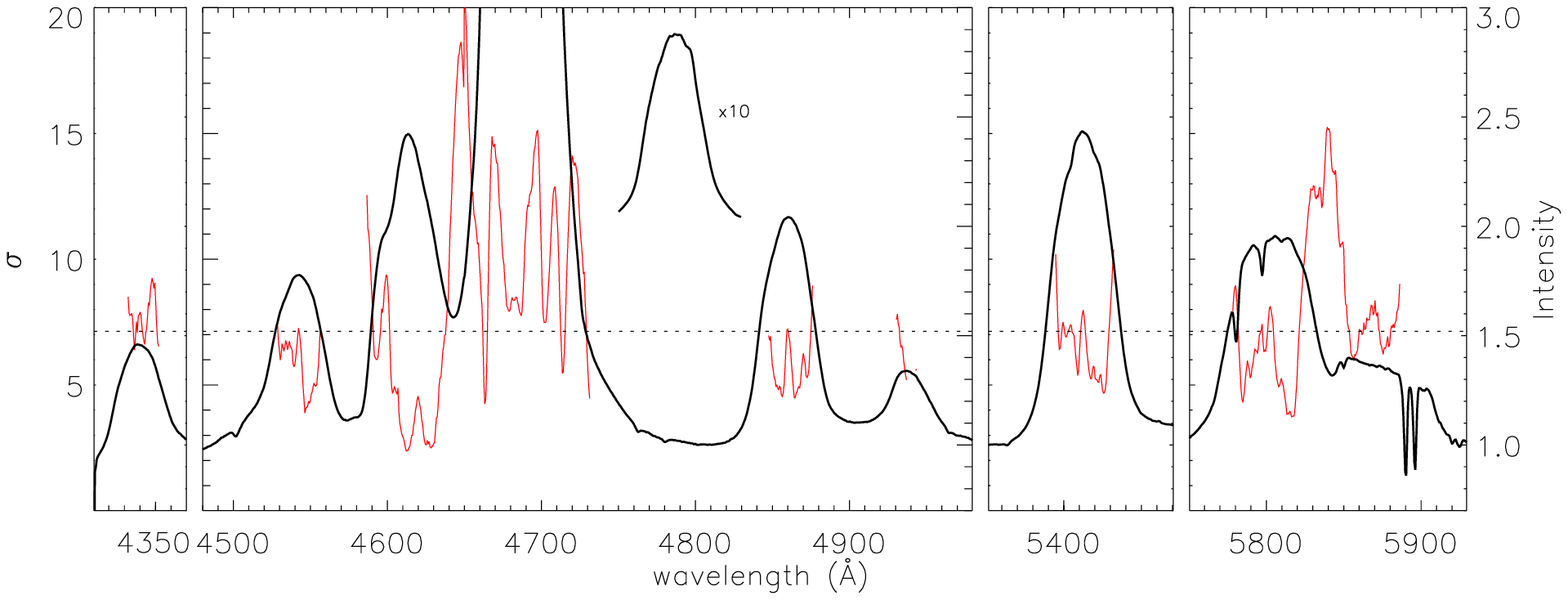}
    \caption{a) Mean spectrum for each of our spectroscopic data runs
      (thick black lines). The $\Sigma$ spectrum, calculated with the
      method introduced by \citet{Ful96} and described in text, is
      overplotted for each run (thin blue lines). Line identification
      is provided in the panel for run 1 and the regions adopted for
      the rectification of the spectra are shown in the panel for run
      5. b) $\sigma$ spectrum as defined in \citet{Stlouis} calculated
      in the wavelength regions where the strongest emission lines are
      present. The dotted line represent the mean $\sigma$-value for
      all emission lines across the observed spectrum.}
    \label{tvs}
\end{figure}

In Figure~\ref{tvs}a, we present the mean spectrum for each of our
spectroscopic data runs listed in Table~\ref{resObs} (thick black
lines). Line identification is provided in the panel for run 1.

To illustrate which parts of our spectra are variable, we have
calculated the temporal variance spectrum (TVS) according to the
method introduced by \citet{Ful96}. Then, assuming that our data are
governed by a reduced $\chi^2$ distribution with N-1 degrees of
freedon, we calculated the $\Sigma$ spectrum as follows~:
\begin{eqnarray}
\Sigma_j(99\%)=\sqrt{\frac{(TVS)_j}{\sigma_0^2\chi_{N-1}^2(99\%)}}
\end{eqnarray}
where N is the total number of spectra and $\sigma_0$ a standardized
dispersion. The $\Sigma$ spectrum for each run is overplotted (in
thin blue line) on the mean spectra in Figure~\ref{tvs}a. A
$\Sigma(99\%)$ value of 3, for example, means that we are 99\%
confident that that part of the spectrum is variable at a 3$\sigma$
level \citep[see ][ for more details]{Stlouis}. It can be easily seen
that all lines show variability at high sigma levels with a high
degree of confidence for each run. Note that since the TVS compares of
the variability level at each pixel with the variability calculated in
the continuum regions, no changes can be identified in the continuum
with this method.

In Figure~\ref{tvs}b, we present the $\sigma$ spectrum which we
defined in \citet{Stlouis}. This spectrum gives the fraction of the
line flux that is variable (and therefore is only defined within
spectral lines). We present 4 regions of our wavelength range which
include spectral lines. Note that each region is not covered by the
same number of spectra. Hence, some are more noisy such as He{\sc
ii}$\lambda$4686 for which we only have $\sim$20 spectra as opposed to
He{\sc ii}$\lambda$5411 for which we have $\sim$320. The reason we
have so little spectra for the He{\sc ii}$\lambda$4686 line is that we
decided to allow this line to saturate in run 4, the run with the
largest number of spectra, in order to increase the total number of
spectra we obtained. Indeed, the alternative would have been to divide
the exposure time by a factor 3 and observe three times as often,
which would have led to tripling the amount of time spent reading the
detector. It can be seen that all lines vary typically at a $\sim$7\%
level. This is exactly the same result found by \citet{Stlouis} for
this star, but based only on 5 spectra.

\subsection{Variability Pattern of Different Spectral Lines}

\begin{figure}[ht]
  \plotone{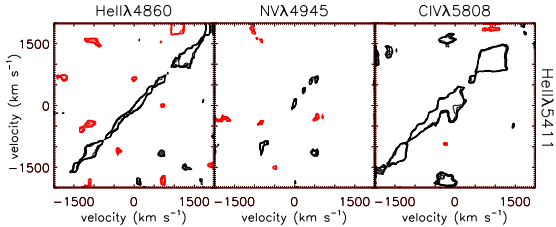}
  \caption{Contour maps of the correlation matrices of He{\sc
  ii}$\lambda$5411 with He{\sc ii}$\lambda$4860, N{\sc v}$\lambda$4945
  and C{\sc iv}$\lambda$5808. The line contours indicate a correlation
  (in black) or an anti-correlation (in red) in the variability
  pattern of the lines. The lowest contour level represents a
  correlation at the 99.5\% confidence level.}
  \label{Spearman}
\end{figure}

In order to compare the variability pattern of different emission
lines, we have calculated correlation coefficients between each
velocity bin of the different emission profiles, using the Spearman
rank-order correlation. This procedure yields a matrix of correlation
that assesses how well the variability pattern of two lines
correlate. For lines with overlapping formation regions (which if
often the case), a correlation should be found if an overdensity is
present in the wind. In the case of line-formation zones far-removed
from the acceleration region, large sections of the wind are
associated with a narrow velocity interval. In such a case, a
correlation would support ``solid'' rotating CIRs. In the case of
perfectly correlated variations, a matrix unity is obtained. In
Figure~\ref{Spearman}, we present the correlation matrices of He{\sc
ii}$\lambda$5411 with He{\sc ii}$\lambda$4860, N{\sc v}$\lambda$4945
and C{\sc iv}$\lambda$5808. The lowest level represents a significant
correlation at the 99.5\% confidence level. The correlation matrix of
He{\sc ii}$\lambda$5411 with He{\sc ii}$\lambda$4686 is not shown
here, but has already been published by \citet{Morel99b}. The
variability pattern of He{\sc ii}$\lambda$5411 is clearly correlated
with that of all other emission lines, except for He{\sc
i}$\lambda$5876 and for N{\sc v}$\lambda$4603/20, which is blended
with the highly variable absorption part of the P~Cygni profile of
He{\sc ii}$\lambda$4686. One can note that, contrarily to
\citet{Morel99b}, we were able to detect a clear correlation between
the variability pattern of the He{\sc ii}$\lambda$5411 and N{\sc
v}$\lambda$4945 lines. This is most likely due to the higher
signal-to-noise ratio and the greater number of spectra in our
dataset.

\begin{figure}[htbp]
    \plotone{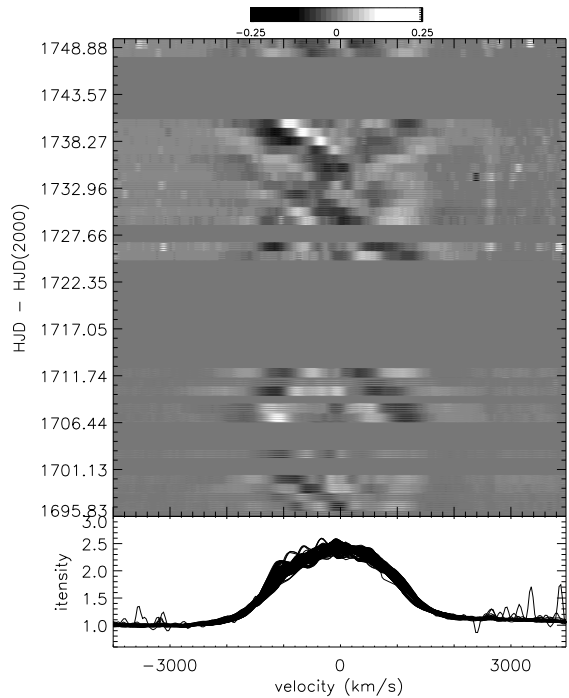}
    \caption{{\it Top panel}: Grayscale of residuals for the He{\sc
    ii}$\lambda$5411 line obtained by subtracting the mean spectrum of all runs from the rectified spectra
    taken during runs 4 and 5. {\it
    Bottom panel}: Superposition of all spectra taken during the two
    runs. Note that for clarity, the time corresponding to daylight
    has been filled by the duplicate of the spectra obtained during
    the previous night.}
    \label{grayWR1}
\end{figure}

In view of the above results, we only show here the variability
pattern of the He{\sc ii}$\lambda$5411 line which is a strong and
isolated line and is common to all our observing runs. The variability
of that line when observed during individual nights is typical of what
has already been reported by \citet{Nieda} for WR\,1, i.e. low level
variations occuring slowly and systematically during the
night. However, the changes are more interesting when displayed in the
form of a grayscale plot covering a time period of more than a week,
as shown in Figure~\ref{grayWR1}, where we plot the complete dataset
of our runs 4 and 5. For clarity, the spectra obtained during each
night have been repeated once to fill the space in the grayscale plot
corresponding to daylight, when no observations were possible. This
way, no free space has been left between two consecutive nights,which
makes it easier to see the structures moving on a timescale of several
days. Although our time coeverage is not ideal, one can begin to
distinguish the typical ``S-type'' pattern caracteristic of CIR-type
variability.

\subsection{Radial Velocity, EW, Skewness and Kurtosis Variations}

\begin{figure}[ht]
  \plotone{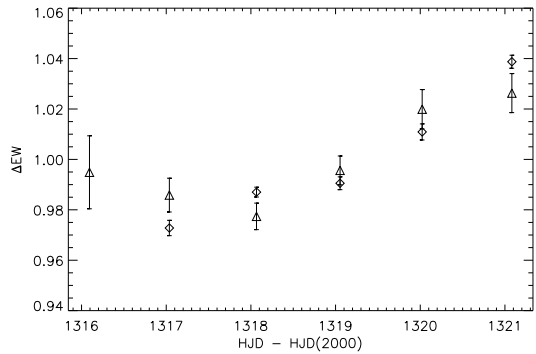}
  \caption{$\Delta$EW of the He{\sc ii}$\lambda$4686 (diamonds) and
He{\sc ii}$\lambda$5411 (triangles) lines observed in run 2. The error
bars are 3~$\sigma$ long. Note that due to probems with the first
spectrum of that run at $\lambda <$5200\,\AA, $\Delta$EW could not be
computed for He{\sc ii}$\lambda$4686 on the first night.}
  \label{CompEW}
\end{figure}

\begin{figure}[ht]
  \plotone{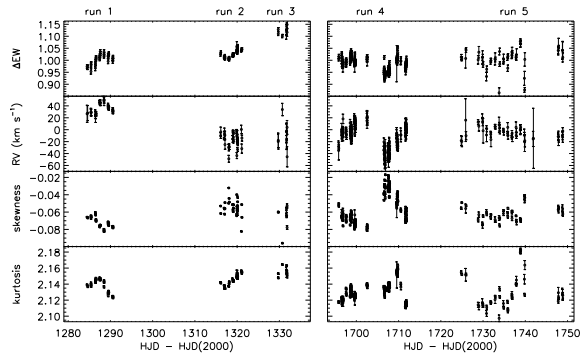}
  \caption{{\it Top panel}: $\Delta$EW of the He{\sc ii}$\lambda$5411
  line, obtained by integrating the line flux and dividing the result
  by the mean value of all the EWs. {\it Second panel from top}: RVs
  obtained by cross-correlating the spectra in the 5140-5980\,\AA\,
  range. {\it Second panel from bottom}: Skewness of the He{\sc
  ii}$\lambda$5411 line calculated as skewness=$\mu_3/\mu_2^{3/2}$,
  where $\mu_n$ is the central moment of order n of the line. {\it
  Bottom panel}: kurtosis of the He{\sc ii}$\lambda$5411 line
  calculated as kurtosis=$\mu_4/\mu_2^2$.}
  \label{Moments}
\end{figure}

We next investigate the variability of the RV of spectral lines as
well as the changes in the global intensity of the line profile by
measuring the equivalent width (EW), the skewness and the kurtosis of
the He{\sc ii}$\lambda$5411 line. The radial velocities were obtained
by cross correlating individual spectra in the wavelength interval
$\Delta \lambda$=5140-5980\,\AA\, with the mean spectra. The RVs of
the different campaigns were obtained separately. We corrected the
systematic shifts between runs by cross-correlating the mean spectra
of the runs. The typical values for the applied corrections ranged
between -20~km~s$^{-1}$ and -40~km~s$^{-1}$. The EWs of emission lines
were calculated by integrating the function $(1-F_\lambda)$, where
$F_\lambda$ is the rectified line flux between $\Delta
\lambda$=5348--5460\,\AA. In order to highlight the variability, we
prefer to use $\Delta$EWs obtained by dividing the EWs by the mean
value of all EWs. In order to verify that the $\Delta$EW of the He{\sc
ii}$\lambda$5411 line is representative of the curve for any other
emission line in the wavelength range of our observations, we compared
its $\Delta$EW to the one of the N{\sc v}$\lambda\lambda$4603/20,
He{\sc ii}$\lambda$4686, He{\sc ii}$\lambda$4860, C{\sc
iv}$\lambda$5808 and He{\sc i}$\lambda$5876 lines. We present in
Figure~\ref{CompEW} the $\Delta$EW of the He{\sc ii}$\lambda$4686 and
He{\sc ii}$\lambda$5411 lines observed in run 2. The changes clearly
follow the same pattern. In general, except for the blended N{\sc
v}$\lambda\lambda$4603/20 and the weakly varing He{\sc i}$\lambda$5876
lines, with the quality of the data in hand, there is no significant
difference that can be seen in the $\Delta$EW value of all observed
lines. The other moments of the He{\sc ii}$\lambda$5411 line are
calculated in the wavelength interval
$\Delta\lambda$=5348-5460\,\AA. The $n^{th}$ central moment is defined
as follows:
\begin{eqnarray}
\mu_n=\Sigma_j(\lambda_j-\bar{\lambda})^n I_j/\Sigma_jI_j
\end{eqnarray}
where
\begin{eqnarray}
\bar{\lambda}=\Sigma_j\lambda_jI_j/\Sigma_jI_j,
\end{eqnarray}
with $I_j$ being the intensity of the line and $\lambda_j$ the
wavelength. The skewness is $\mu_3/\mu_2^{3/2}$ and the kurtosis is
$\mu_4/\mu_2^{2}$. All these quantities are plotted as a function of
time in Figure~\ref{Moments}.

On average, the EW of the He{\sc ii}$\lambda$5411 line increased by
$\sim$15~\% during the first three runs ({\it left panel}) and by
$\sim$5~\% during the last two ({\it right panel}). The RVs show
long-term changes with a full amplitude K$\sim$70~km~s$^{-1}$ as well
as shorter timescale changes which are particularly apparent during
run 4. However, the skewness, which characterizes the asymmetry of the
line profile, is strongly anti-correlated with the RVs. Indeed, these
values are anti-correlated with a 90~\% confidence level when we
remove run 5, which seems to have abnormally high dispersion. This
strong anti-correlation indicates that the RV changes that we measure
essentially come from changes in the degree of asymmetry of the line
profile of He{\sc ii}$\lambda$5411, which dominates the spectrum in
the wavelength interval in which the cross-correlation was carried
out. Consequently, if WR\,1 shows intrinsic RV variability in He{\sc
ii}$\lambda$5411, it must be at a very small level and therefore is
difficult to detect in view of the line-profile variability. Both EW
and RV of the He{\sc ii}$\lambda$5411 line vary with an amplitude of
at most $\sim$5~\% in a single night. This can be compared with the
observations of \citet{Nieda} who found EW variations of He{\sc
ii}$\lambda$5411 of $\sim$15\% for their first run with a structured
variability pattern over a $\Delta t$=5 day period, but only $\sim$5\%
for their second run on a timescale of $\sim$2~days. Unfortunately,
\citet{Flor} did not observed this line. The variations in EW are
clearly epoch-dependant.

\subsection{Search for Periods}

\begin{figure}[htbp]
  \plotone{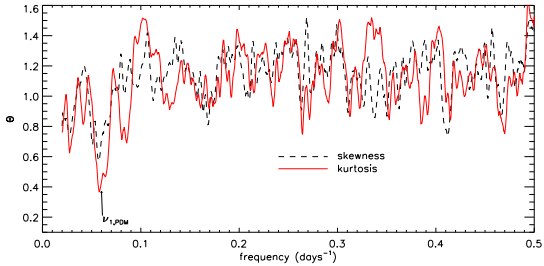}
  \caption{$\Theta$-spectrum obtained from the PDM analysis of the
    skewness curve (black, dashed line) and the kurtosis curve (red,
    solid line) of He{\sc ii}$\lambda$5411, presented in
    Figure~\ref{Moments}. The photometric frequency $\nu_{1,PDM}$
    discussed in Section~\ref{resPhot} is indicated with an arrow.}
  \label{permom}
\end{figure}

\begin{figure}[htbp]
  \plotone{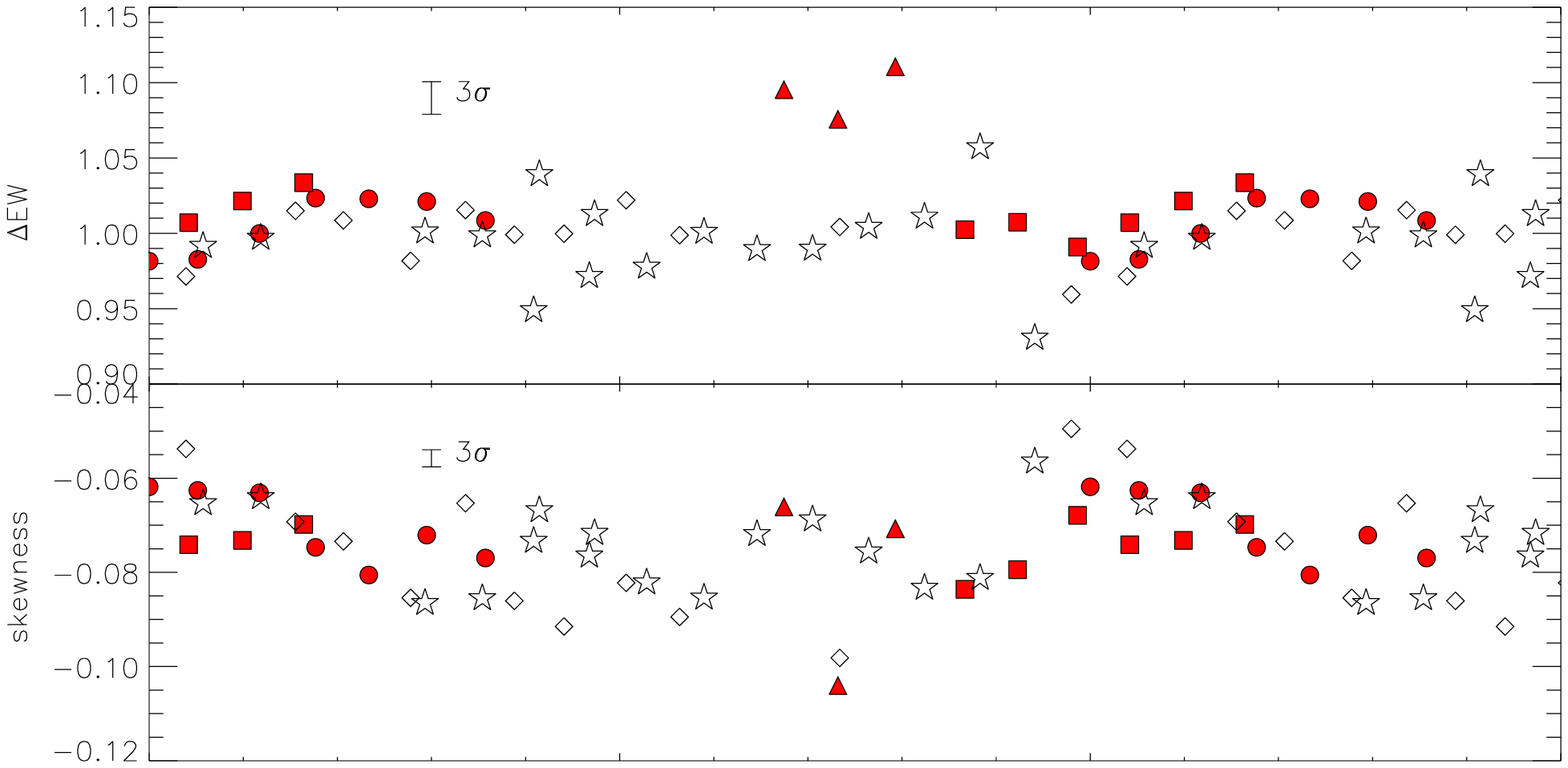}
  \plotone{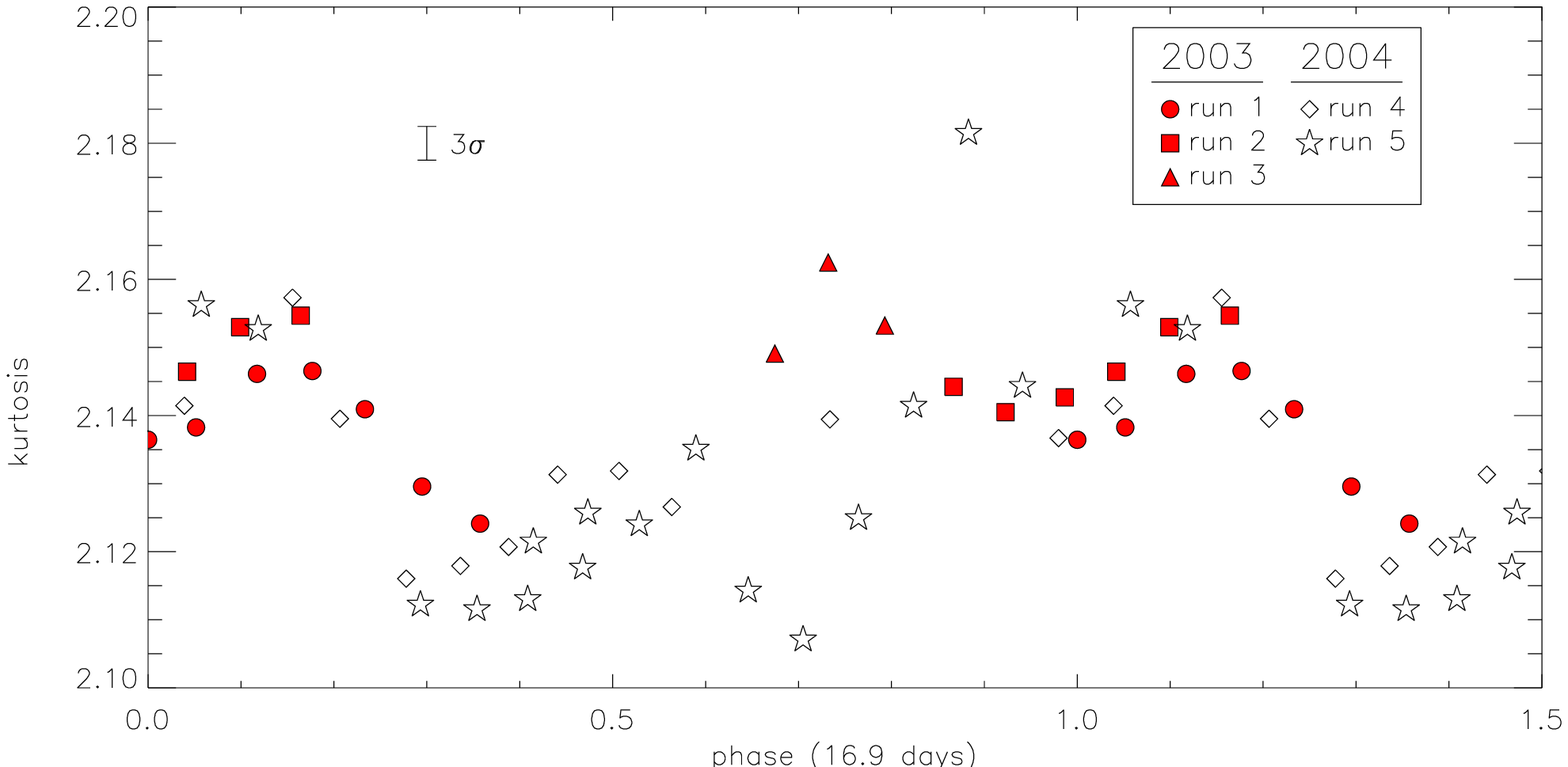}
  \caption{The $\Delta$EW, skewness and kurtosis curves for He{\sc
  ii}$\lambda$5411 folded with the frequency
  $\nu_{1,PDM}$=0.059~days$^{-1}$. Each point is the mean value for a
  single night. Run 1 is represented by circles, run 2 by squares, run
  3 by triangles, run 4 by diamonds and run 5 by stars.}
  \label{folmom}
\end{figure}

A period search was performed using the PDM method on the EWs,
skewness and kurtosis variations (but not on the RVs, due to these
similarity with the skewness). No clear period could be found in the
EW curves. However, the periodograms calculated for both the skewness
and the kurtosis curves have a significant peak at 0.059~day$^{-1}$,
i.e. the photometric frequency $\nu_{1,PDM}$ discussed in the previous
section (see Figure~\ref{permom}). The curves folded with that
frequency are shown in Figure~\ref{folmom}. Each point is the mean
value for a single night. Run 1 is represented by circles, run 2 by
squares, run 3 by triangles, run 4 by diamonds and run 5 by stars. The
folded curves for the $\Delta$EW and the skewness have quite a large
scatter and poorly confirm the detected period. However, the folded
curve for the kurtosis is more promising. The largest scatter is found
between phases 0.55 and 1.0, but it can be greatly reduced if one
conciders separately runs 1, 2 and 3 observed in 2003 ($\Delta t\sim
50$~days) and runs 4 and 5 observed in 2004 ($\sim
55$~days). Interestingly, this is also the time interval aftehr which
a change in the shape of our lightcurve was observed. One maximum,
centered at phase$\sim$0.1 is common to all runs. The variations
between phases 0.55 and 1.0 can be described as one additionnal
maximum for runs 1, 2 and 3 and a minimum (around phase 0.6) followed
by a maximum (around phase 0.9) for runs 4 and 5.


It is not necessarily trivial to link the variability of the global
moments of the lines and the wind variability of the star. Here, we
present an investigation that aims to reveal the periodicity of the
movement of the extra bumps observed superposed on the emission line
profiles. These can be seen moving from one side of the profile to the
other for runs 4 and 5 in Figure~\ref{grayWR1}. We define the scalar
value $\sigma_{res}$, i.e. the standard deviation of the residuals
obtained from the subtraction of two spectra observed separately with
a time interval of $\Delta t$. Theoretically, when two spectra are
obtained at the same point of a given variability pattern and
therefore should be most similar, $\sigma_{res}$ reaches a minimum and
takes the value of the quadratic sum of the noise levels of both
spectra. Also, any significant difference between the two spectra will
increase the value of $\sigma_{res}$ proportionally to the magnitude
of that difference. The great advantage of $\sigma_{res}$ is that its
value depends weakly on the epoch depedency of the signal. Indeed,
even if the signal changes after 4 or 5 cycles (see
Section~\ref{resPhot}), all spectra from any cycle can be used at a
given $\Delta t$, since only the spectra separated by the time
interval $\Delta t$ are used. The effect of the epoch dependency
becomes significant for $\Delta t$ greater than the coherence time,
i.e. in this case, when two spectra separated by more than (at least)
$\sim$4 cycles ($\sim$67~days) are compared.

\begin{figure}[htbp]
  \plotone{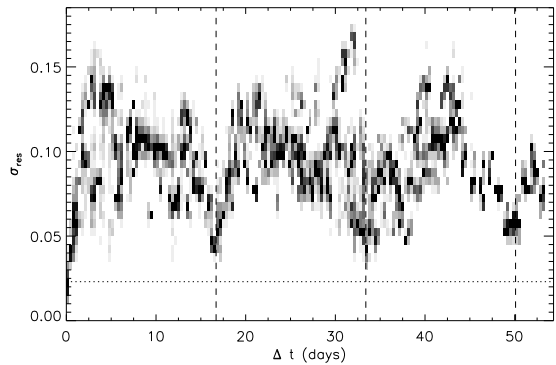}
  \caption{The value of $\sigma_{res}$ measured for the He{\sc
  ii}$\lambda$5411 line as a function of $\Delta t$. $\sigma_{res}$ is
  defined as the standard deviation of the residual obtained from the
  subtraction of two spectra observed separately with a time interval
  of $\Delta t$. The dotted line indicates the minimum value for
  $\sigma_{res}$ due to the noise level. The dashed lines mark $\Delta
  t$=n$\times$16.9 where n=1, 2 and 3. The vertical bar in the top
  left corner indicates the 1 sigma scatter of the $\sigma_{res}$
  value for a given $\Delta t$.}
  \label{sigvsT}
\end{figure}

In Figure~\ref{sigvsT} we plot the values of $\sigma_{res}$ measured
within a wavelength interval of 65\,\AA\, centered on
$\lambda$=5411\,\AA\, for $\Delta t < 55$~days, using all the spectra
from our 5 observing runs. The wavelength interval has been chosen to
include the part of the He{\sc ii}$\lambda$5411 line where most of the
lpv occurs, without adding too much of the less intense parts that do
not contain much information and have a much lower SNR. For clarity,
we have organized the values of $\sigma_{res}$ as a function of
$\Delta$t in a grayscale. In black are plotted the regions with the
higher sumber of data points and, in white, the regions with no data
point. Since the spectrum of WR\,1 does not vary very much within a
single night (see above), we assume that the mean value of
$\sigma_{res}$ measured for all $\Delta t < 1$~day can be used to
determine the minimum value for $\sigma_{res}$ (plotted with a
dotted-line). Also, the spectra have been normalized in amplitude to
avoid any change coming from the line dilution caused by the variation
in the continuum that we observe in photometry.

\begin{figure}[ht]
  \plotone{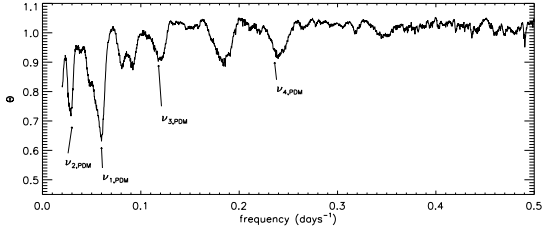}
  \caption{$\Theta$-spectrum obtained from the PDM analysis of the
    $\sigma_{res}$ curve. The photometric frequencies $\nu_{1,PDM}$,
    $\nu_{2,PDM}$, $\nu_{3,PDM}$ and $\nu_{4,PDM}$ discussed in
    Section~\ref{resPhot} are indicated by arrows.}
  \label{persigvsT}
\end{figure}

The periodogram calculated for the $\sigma_{res}$ vs $\Delta t$ curve
using the PDM method is shown in Figure~\ref{persigvsT}. The four
frequencies $\nu_{1,PDM}$, $\nu_{2,PDM}$, $\nu_{3,PDM}$ and
$\nu_{4,PDM}$ are indicated by arrows. In Figure~\ref{sigvsT}, a
vertical dashed-line is plotted at a constant interval of
16.9~days. Interestingly, $\sigma_{res}$ reaches a minimum value near
$\Delta t$=n$\times$16.9~days, where n=1, 2 and 3. These minima are
significant when compared with $\sigma_{1n}$, but they do not reach
the minimum value corresponding to the noise level. This may indicate
that even after one cycle, the variability pattern has changed
slightly (but not drastically). Two other small minima can be seen at
$\Delta t \sim$4.5 and 11.8~days, but they do not repeat at greater
$\Delta t$. This can be caused by line-profile variability that shows
a similar pattern occuring more than once within the period for at
least one cycle (as seen in the photometry).

\begin{figure}[htbp]
\plottwo{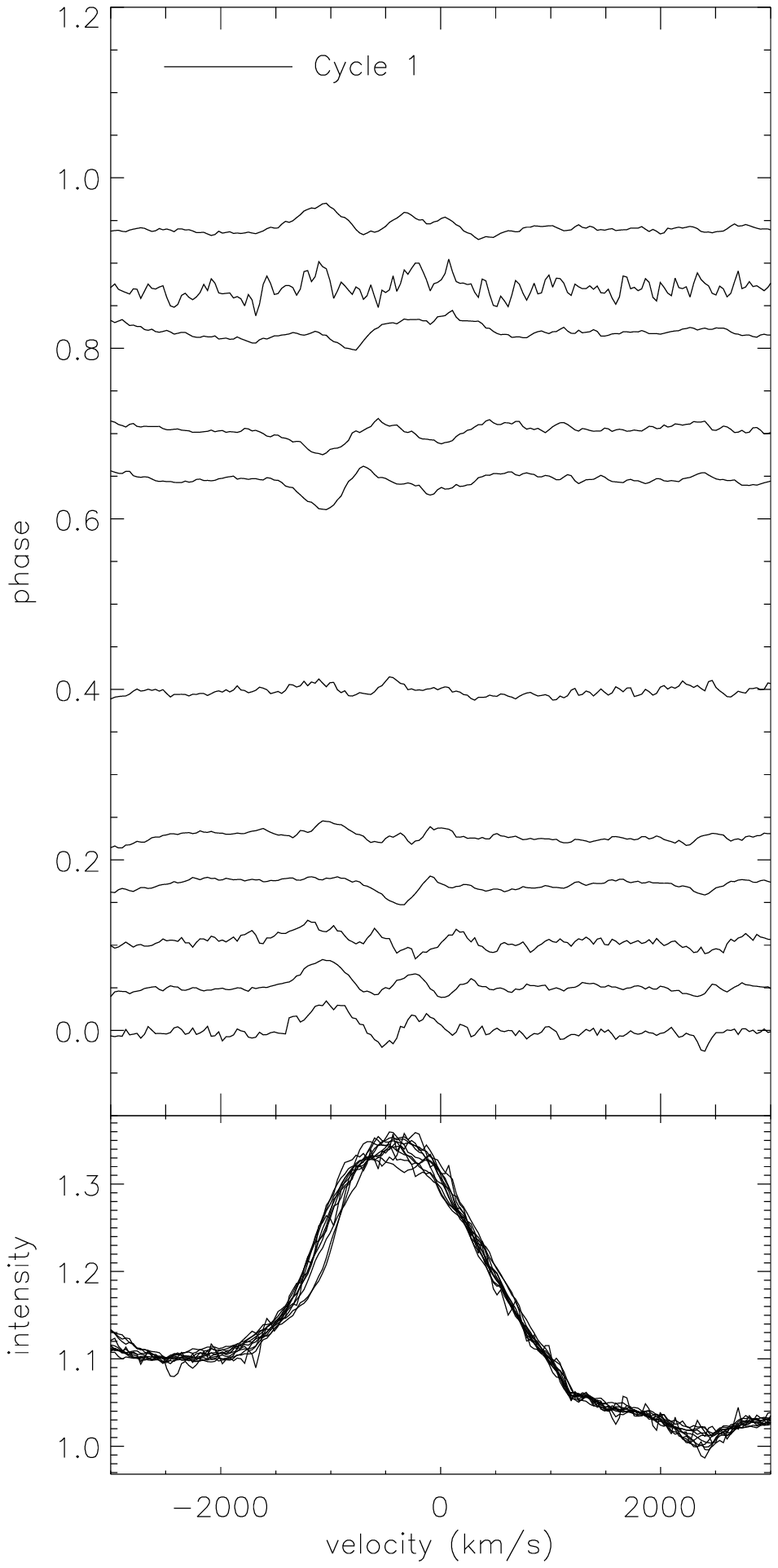}{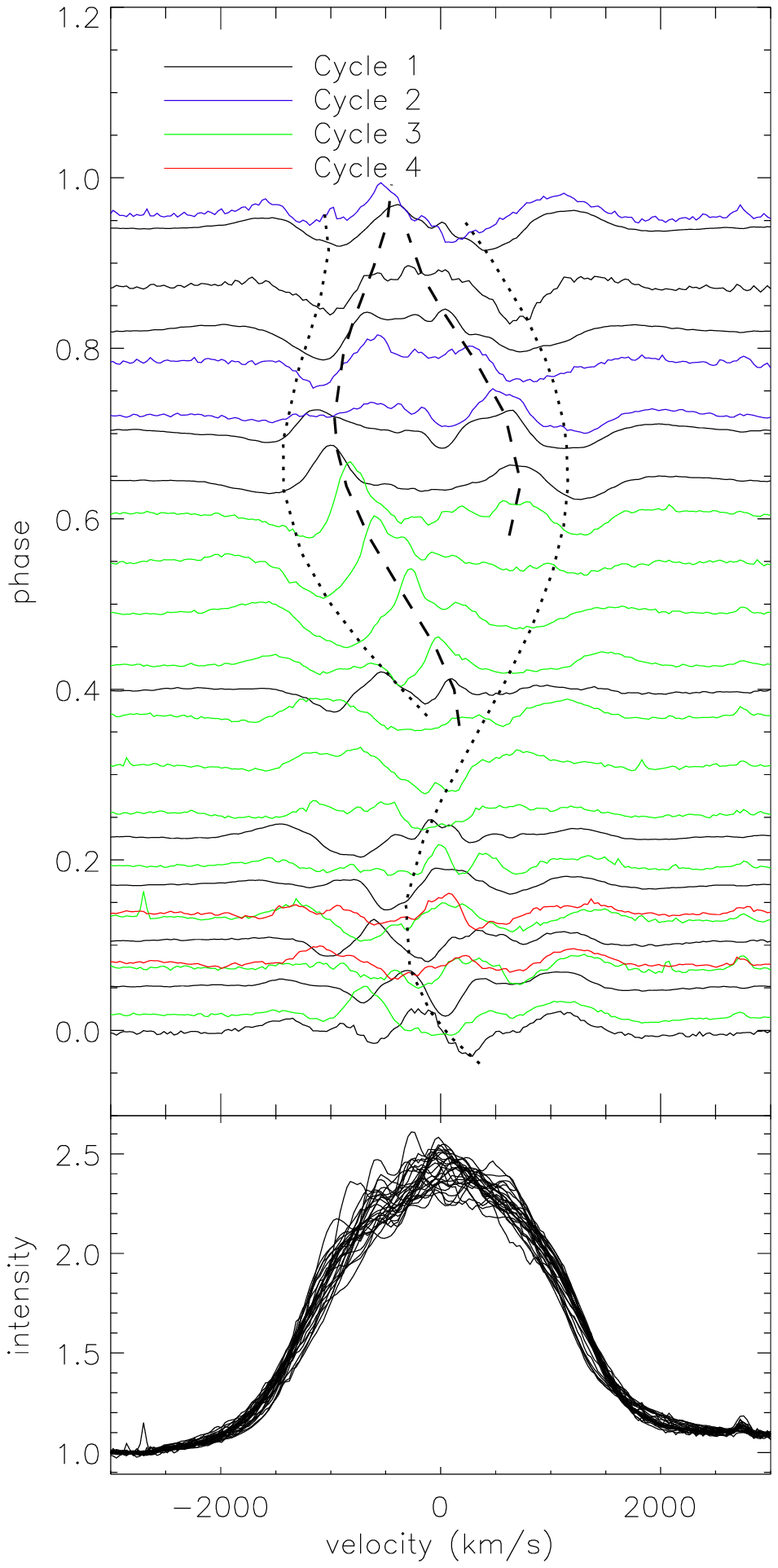}
%  \begin{minipage}[t]{.45\textwidth}
%    \plotone{fig13a.eps}
%  \end{minipage}
%  \begin{minipage}[t]{.45\textwidth}
%    \plotone{fig13b.eps}
%  \end{minipage}
  \caption{Residual of nightly mean spectra of runs 4 and 5 folded
  with a period of 16.9~days ({\it left panel}:~N{\sc v}$\lambda$4945;
  {\it right panel}:~He{\sc ii}$\lambda$5411). In black we plot the
  residuals from the first cycle covered, in blue the second, in green
  the third and in red the fourth. Moving structures can be followed
  and are shown by dashed (for bumps) and dotted lines (for dips). In
  the {\it bottom} panel, we plot individual intensity spectra
  superposed.}
  \label{montage}
\end{figure}

Finaly, in order to illustrate the line-profile variability pattern,
we display in Figure~\ref{montage} the residuals obtained by
subtracting the mean spectrum of all runs from the mean spectra of
each night of runs~4 and 5 for the N{\sc v}$\lambda$4945 and He{\sc
ii}$\lambda$5411 lines (note that the N{\sc v}$\lambda$4945 line was
not observed during run~5). The location on the y-axis of the residual
spectra is determined by the phase (P=16.9~days) at which the spectrum
was taken. Different colors are assigned to residuals from different
cycles.

One can see from Figure~\ref{montage} that the residuals at nearby
phases are not always strictly identical (as expected from the above
analysis). However, some features seem to survive for several
cycles. The clearest case is that of the bump that appears at $v
\sim$+250~km~s$^{-1}$ at phase $\phi$=0.40 of the first cycle
(HJD-HJD(2000)=1702.54), that moves toward negative velocities during
phases $\phi$=0.44-0.61 of the third cycle
(HJD-HJD(2000)=1736.85-1739.83), and continues its blueward motion
during phases $\phi$=0.65-0.70 of the first cycle
(HJD-HJD(2000)=1706.70-1707.70) before disappearing. More moving
structures can be followed on Figure~\ref{montage} and are shown by
dashed (for bumps) and dotted lines (for dips). One can note that the
motion of the bumps explains the changes we have observed in the
central moments of the He{\sc ii}$\lambda$5411 line presented
above. Indeed, the two biggest bumps trace a sinusoidal trajectory on
the line, both in opposite directions. Hence, it is expected that the
skewnesss, which measures the asymmetry of the line, would not vary
too much due to the motion of the bumps, and that the kurtosis, which
measures the degree of ``peakedness'' of the line, would be more
sensitive to it.

\section{Discussion}\label{dis}

\citet{Stlouis} have already shown that WR\,1's spectrum shows
large-scale lpv with a similar amplitude to what is observed for WR\,6
and WR\,134, two presumably single WR stars showing periodic
photometric and spectroscopic variability
\citep{Morel97,Morel98,Morel99a,Morel99b}. With the intensive
monitoring campaign presented here, we are now able to conclude that,
as for those two stars, WR\,1 presents a unique period in its
epoch-dependent light-curve and spectral varibility. The changes
consist in large-scale bumps moving periodically from one side of
emission lines to the other. The origin of this type of variability is
debated in several papers and the two main interpretations put forward
are a binary system (with a compact or low-mass companion) and a
rotating non-spherically symmetric wind. In what follows, we discuss
these two possible interpretations in the context of the variability
of WR\,1 as characterized in this work, as well as that of WR\,6 and
WR\,134 as detailed in the literature. We also discuss the variability
of the WR star member of the binary WR\,137, which shows large-scale
lpv with a 1.2d-period that cannot be caused by the O9 companion nor
the wind-wind collision zone, since the orbital period is
$\sim$13~years and the average distance between the two components is
high.

\subsection{The Binarity Scenario}

Like WR\,6, WR\,134 and WR\,137, WR\,1 shows very small RV amplitude,
consistent with null, and its soft X-ray luminosity is within the
normal range for single WR stars \citep[see ][]{Morel99b}. Hence, it
is completely excluded that its companion is an OB star. The presence
of a compact companion, such as a neutron star or a black hole, in
WR\,1, WR\,6 and WR\,134 was not seen as a viable interpretation by
\citet{Ign2}, \citet{Morel97}, \citet{Skinner} and
\citet{Morel99a}. Indeed, the strong correlation between the lpv of
He{\sc ii} lines with that of a highly ionized line such as N{\sc
v}$\lambda$4945 indicates that the ionizing shell around the compact
companion would have to be extremely large and would emit a high X-ray
flux, which is not observed \citep{Morel97,Morel99a,Pol}.

Therefore the only remaining possibility for the binary scenario for
WR\,1, WR\,6 and WR\,134 is that of a low-mass, non-compact
companion. Indeed, assuming circular orbits and taking the published
RV variation amplitudes measured for the three stars \citep[][ and
this work]{Morel97,Morel98} as the maximum possible value (recall that
RV variations are highly dominated by lpv) the periods lead to an
upper limit for the mass of such a companion of 8~M$_\odot$ in all
cases, if the inclination of the orbital plane is higher than
20$^\circ$. As for WR\,137, the scatter in RV on a timescale of a few
days is $\sim$10~km~s$^{-1}$ \citep{Lef} and its measured
M~sin$^3$i=3.4$\pm$1.0~M$_\odot$. Hence, if the inclination of the
orbital plane is higher than 20$^\circ$, the mass of the companion
would have to be as small as 0.8~M$_\odot$ in an orbit of
7.6~R$_\odot$, which is very unlikely. But, can a low-mass companion
be responsible for such a complex and epoch-dependant pattern of
photometric and spectroscopic variability?  \citet{More1} and
\citet{More2} have shown that tidal interactions involving a
relatively low-mass companion in a binary system can produce very
small-scale surface oscillations leading to lpv in photospheric
absorption features. However, it is unclear how such a process can
affect the massive wind of a WR star and lead to large-scale
variability of the strong emission lines. It is possible that it could
serve as a seed mechanism for CIRs. However, in such a case, it is
likely that the period that would be detected would be the rotation
period of the WR star rather than the orbital period of the binary, if
not a combination of both.

Therefore, although very unlikely, it is not yet possible from the
observations to definitively exclude the possibility that WR\,1, WR\,6
and WR\,134 are binaries, and that WR\,137 has a third low-mass
companion. However, the only viable binary scenario involves the
formation of massive binaries with a high initial-mass ratio (up to
10), which seems to be extremely unlikely both observationally and
theoretically \citep[e.g.][]{Gar,Kob,Bon}.

\subsection{The Non-Spherically Symmetric Wind Scenario}

If WR\,1, WR\,6, WR\,134 and the WR star in the WR\,137 binary system
are single stars, the periodic variability very likely originates in
an asymmetry in their wind modulated by the stellar
rotation. Spectropolarimetric observations of WR\,6, WR\,134 and
WR\,137 \citep{Schulteladbeck91,Robert92,Harries98} have shown the
presence of an intrinsic continuum polarization component due to
electron scattering, indicating that the winds of these stars are not
spherically symmetric. This is revealed by the depolarization of
emission lines compared to the neighboring continuum. It was also
found that the degree of depolarisation of spectral lines increases
with decreasing ionisation state.  This is interpreted to be a
consequence of the ionisation stratification of hot stellar winds; as
scattering lines with a higher degree of ionization are formed deeper
in the wind, their polarisation level is closer to that of the
continuum because of the higher density close to the core and
therefore their level of depolarisation is lower. If a line has a
strong recombination component, wich is not polarised, its
polarization level will be even lower and the depolarization with
repect to the continuum stronger. Finally, the level of linear
polarization of the continuum as measured in braodband light of WR\,6,
WR\,134 and WR\,137 is extremely variable and periodic
\citep{Dri,Mof93}. This suggests that the wind density has a varying
asymmetric distribution, such as density structures that extend rather
far in the wind. Unfortunately, there is no published high
signal-to-noise spectropolarimetric nor continuum polarization
observation of WR\,1. The only observation so far was carried out by
\citet{Schmidt88} and does not show a depolarization in its emission
lines compared to the underlying continuum. The latter, however, shows
a significantly high level of polarization, although it remains to be
demonstrated that this is intrinsic to the star rather than
interstellar in origin. If the polarization is proved to be
interstellar, this would explain why no depolarization is observed in
the lines since the continuum is not polarized to start with.  If, on
the other hand, the light from the star is confirmed to be highly
polarized, a possible explanation for the lack of depolarization in
the lines is that for some yet unknown reason, the region in which the
lines arise shows a degree of polarization as high as that in which
the continuum arises. Also, the epoch-dependent nature of the changes
could mean that relatively quiet periods can exist. Therefore, more
spectropolarimetric observations are needed to verify wether the
polarization of the emission lines of WR\,1 varies with time.

The possible existence of large-scale density structures, such as
CIRs, was first proposed by \citet{Mul}. Following this idea,
\citet{Cranmer96} modeled the propagation of CIRs in a hot,
radiatively, line-driven stellar wind. In that context, CIRs are
caused by perturbations at the base of the wind, which in turn could
be caused for example by a magnetic field or pulsations. These
perturbations propagate through the wind while being carried around by
rotation. This generates spiral-like structures in the density
distribution that can lead to a characteristic, large-scale, periodic
variability pattern in WR-wind emission lines, extremely similar to
what is observed in the WR stars we discuss here
\citep{Dessart02}. Thus, taking into account the photometric and
spectroscopic periodic variability, the spectropolarimetric
observations and the low soft and hard X-ray fluxes, we conclude that
CIRs constitute an extremely likely interpretation. As for the
epoch-dependency, it can easily be explained by the variable behaviour
of perturbations that generate CIRs together with a finite lifetime of
the structures in the wind. Indeed, even if the period of the
variability caused by the motion of CIRs is always the same, the
number and position of the CIRs may change depending on conditions at
the surface. The origine of CIRs is still debated and no clear
theoritical predictions of the typical lifetime of a CIR have been
made yet. However, one can speculate that it will depend on the
lifetime of the perturbation at the base of the wind and, to a lesser
extent, on the flow speed in the wind. Observationaly, it could be
determined from continuous observations during a great number of
rotational periods (e.g. several weeks for WR\,6 and WR\,134).

\subsection{Remarks on the putative CIRs in the wind of WR\,1}\label{con}

We believe that the best scenario to explain the photometric and
spectroscopic variability we have detected in WR\,1 is the presence in
its wind of large-scale structures, most likely CIRs. In that case,
each CIR would translate into a bump in the light-curve and a bump
over the spectral emission lines.

In theory, CIRs do not suffer from differential rotation and if the
recurrence of the changes is caused solely by their rotation, they
provide a direct measurement of the rotation period of the underlying
star at the position in which they originate, most likely close to the
stellar surface. Thus, assuming a value for the radius of WR\,1 of
2.2~R$_\odot$ \citep[a radius that corresponds to a Rosseland optical
depth of 20;][]{Ham3}, we obtain an equatorial rotational velocity of
6.5~km~s$^{-1}$. This is an order of magnitude lower than the values
obtained for WR\,6 (40~km~s$^{-1}$) and WR\,134 (70~km~s$^{-1}$) when
assuming a radius of R$_\ast \sim 3 R_\odot$ \citep{Ham3}. However,
all these rotational velocities are in agreement with the very small
values predicted for WR stars by massive-star evolutionary models at
solar metalicity \citep{Mey}. The order of magnitude difference
between the rotation velocity of WR\,1 and that of WR\,6 and WR\,134
is not completely surprising since the observed rotational velocity of
a massive star at the WR evolutionary stage depends on several
parameters such as the initial stellar mass, the mass-loss rate and
the age of the star (time spent as a WR star). Interestingly, adopting
a radius of 4.5~R$_\odot$ \citep[corresponding to the radius of the
hydrostatic core; ][]{Nug} for WR\,137, we deduce an equatorial
rotational velocity of 275~km~s$^{-1}$, i.e. more than half the
breakup velocity. If confirmed, such a fast rotation for a WC star
would render WR\,137 an interesting candidate for an eventual
long-term gamma-ray burst.

Inspired by the analysis of \citet{Dessart02}, we can estimate the
inclination of the CIRs in the wind of WR\,1. Indeed, the maximum
Doppler velocity that a bump associated with a CIR reaches during its
motion on an emission line is $v_{max}=\pm v_{lfr} \rm{cos}(\theta)$,
where $v_{lfr}$ is the velocity of the wind at the radii corresponding
to the formation region of the observed emission line over which the
bump is observed, and $\theta$ is the inclination angle of the CIR
with the line-of-sight. In Figure~\ref{montage}, we track two bumps
that reach $v_{max1}\sim$ --1300~km~s$^{-1}$ and $v_{max2}\sim$
+700~km~s$^{-1}$ on the He{\sc ii}$\lambda$5411 line. Assuming that
the velocity law of the wind can be described by a $\beta$-law for 
which the velocity as a function of radius can be written as follows
\citep{Castor}:
\begin{eqnarray}
v(r)=v_\infty\left(1-\frac{R_\ast}{r}\right)^\beta,
\end{eqnarray}
where $v_\infty$ is the terminal velocity of the wind, the value of
$v_{lfr}$ for He{\sc ii}$\lambda$5411 can be estimated using the
emissivity function described by \citet{Lep}. However, that function
depends on the value of $\beta$. Unfortunately, $\beta$ cannot be
estimated observationally for WR\,1, since no clumps are observed in
the spectra. We can, however, give estimates of $\theta$ as a function
of $\beta$. If $\beta$=1, 2 or 3 the first CIR has an inclination
angle of $\theta_1$=$\pm$40$^\circ$, $\pm$38$^\circ$ or $\pm$30$^\circ$,
respectively, and the second bump has $\theta_2$=$\pm$65$^\circ$,
$\pm$64$^\circ$ or $\pm$62$^\circ$, respectively. When $\beta$ is greater than
3, the velocity of the He{\sc ii}$\lambda$5411 line formation region
is lower than 1300~km~s$^{-1}$. Hence, under our first assumptions, we
deduce that the $\beta$ value for the wind of WR\,1 should be lower
than 3.

Of course, the association of the period of the CIRs with the rotation period
of the underlying star remains to be confirmed. In the wind of OB
stars, the presence of CIRs is thought to be at the origin of
periodically recuring Discrete Absorption Components (DACs) in P~Cygni
profile of UV resonance lines \citep[e.g. ][]{Fullerton97}. In many
cases, the period of the outward moving DACs is found to correspond to $v_{eq}
\sin{i}$, where $v_{eq}$ is the rotational velocity at the equator and
$i$ the inclinaison of the rotational axis with the line of sight
\citep{Hen,Prin}. But in the case of HD\,64760, a B0.5 Ib star, the
CIRs rotate more slowly than the stellar surface. \citet{Lob} carried
out a hydrodynamical simulation of the CIRs for that star, following
\citet{Cranmer96}, with the difference that they allowed the ``spots''
at the origine of the CIRs to move on the stellar surface. They were
able to obtain a fairly good fit to the lpv of the Si{\sc
iv}$\lambda\lambda$1394,1403 and concluded that, for that star, the
origin of the CIRs must be the interference pattern of a number of
non-radial pulsations at the surface of the star.

So far, there is no model for the ``spots'' that are at the origin of
the creation of the CIRs and there are only a limited number of
datasets available that could potentially be used to confirm either
the pulsational or the magnetic origin of CIRs. The range of frequency
expected for pulsations in WR stars is not well determined. Up to
date, only one detection of a relatively stable 9.8~h period that can
be attributed either to g-modes \citep{Tow} or stange-mode pulsations
\citep{Dor} has been claimed in WR\,123 \citep{Lef2}. No pulsational
period of a few days is currently known nor predicted. Also, no
detection of a magnetic field has been achieved in a WR star so far;
only an upper limit of $\sim$ 25 Gauss for WR\,6 has been claimed 
\citep{Stl}.

\acknowledgments

We thank Anthony~F.~J. Moffat who greatly contributed in the success
of the photometric campaign. ANC thanks Dominique Ballerau and Jacques
Chauville for their precious help with the observations at OHP and the
spectra extraction. NSL thanks the Natural Sciences and Engineering
Research Council (NSERC) of Canada for financial support. Finally, we
thank the anonymous referee who, through very detailed comments,
contributed to greatly improve the quality of this paper.

{\it Facilities:} \facility{CFHT}, \facility{OHP}, \facility{DAO}, \facility{OMM}.

\end{document}